# Convergence Time Analysis of Quantized Gossip Consensus on Digraphs

Kai Cai and Hideaki Ishii

**Abstract**

We have recently proposed quantized gossip algorithms which solve the consensus and averaging problems on directed graphs with the least restrictive connectivity requirements. In this paper we study the convergence time of these algorithms. To this end, we investigate the shrinking time of the smallest interval that contains all states for the consensus algorithm, and the decay time of a suitable Lyapunov function for the averaging algorithm. The investigation leads us to characterizing the convergence time by the hitting time in certain special Markov chains. We simplify the structures of state transition by considering the special case of complete graphs, where every edge can be activated with an equal probability, and derive polynomial upper bounds on convergence time.

## I. INTRODUCTION

Inspired by aggregate behavior of animal groups and motion coordination of distributed robotic networks, the *consensus* problem has been extensively studied in the recent literature of systems control (e.g., [1]–[3]). The objective of consensus is to have a population of nodes, each possessing an initial state, agree eventually on *some* common value through only local information exchange. This problem is also intimately related to oscillator synchronization [4], load balancing [5], and leader election [6]. The *averaging* problem is of a special form, with the goal to decentrally compute the *average* of all initial states at every node.

We have recently proposed in [7], [8] randomized gossip algorithms which solve the consensus and averaging problems on directed graphs (or digraphs), under a quantization constraint that each node has

The authors are with the Department of Computational Intelligence and Systems Science, Tokyo Institute of Technology, 4259-J2-54, Nagatsuta-cho, Midori-ku, Yokohama 226-8502, Japan. Phone/Fax: +81-45-924-5371. Emails: caikai@sc.dis.titech.ac.jp, ishii@dis.titech.ac.jp. This work was supported in part by the Ministry of Education, Culture, Sports, Science and Technology in Japan under Grant-in-Aid for Scientific Research, No. 21760323.

an integer-valued state. In particular, our derived connectivity condition ensuring average consensus is weaker than those in the literature [2], in the sense that it does not postulate balanced topologies. Here the main difficulty is that the state sum of nodes cannot be preserved during algorithm iterations. This scenario was previously considered in [9], [10], where averaging is guaranteed in expectation but there is in general an error in mean square and with probability one. By contrast, we overcome this difficulty by augmenting the so-called "surplus" variables for individual nodes so as to maintain local records of state updates, thereby ensuring average consensus almost surely.

In this paper and its conference precursor [11], we investigate the performance of our proposed algorithms by providing upper bounds on the *mean convergence time*. The state transition structures resulting from these algorithms turn out to be rather complicated. Hence in our analysis on convergence time, we focus on the special case of complete graphs. The analysis is still challenging, but we will also discuss that the general approach can be useful for other graph topologies. First, for the consensus algorithm, we find that the mean convergence time is $O(n^2)$. To derive this bound, we view reaching consensus as the smallest interval containing all states shrinking its length to zero. This perspective leads us to characterizing convergence time by the *hitting* time in a certain Markov chain, which yields the polynomial bound. Second, we obtain that the mean convergence time of the averaging algorithm is $O(n^3)$. As the original algorithm in [7], [8] is found to induce complex state transition structures, we have suitably revised it to manage the complexity. For the modified algorithm, a Lyapunov function is proposed which measures the distance from average consensus. We then bound convergence time by way of bounding the number of iterations required to decrease the Lyapunov function; the latter is again characterized by the hitting time in a special Markov chain.

Our work is related to [12]–[15], which deal also with the convergence time of gossip averaging algorithms with quantized states. In [12], a Lyapunov approach is adopted and polynomial bounds on convergence time are obtained for fully connected and linear networks. The work [13] generalizes these bounds to arbitrarily connected networks (fixed or switching), utilizing the results on the meeting time of two random walks on graphs. Also, bounds for arbitrarily connected networks are provided in [14], [15]; these bounds are, however, in terms of graph topology rather than the number of nodes. In these cited references, a common feature is that the graphs are undirected. By contrast, our algorithm in [7], [8] is designed for *arbitrary strongly connected* digraphs, and we are interested in studying the corresponding convergence time.

To bound the convergence time, a frequently employed approach is to bound the decay time of some suitable Lyapunov functions [12], [16]. In particular, [16] derives tight polynomial bounds on the convergence time of synchronized averaging algorithms, with either real or quantized states. In addition, [17] investigates a variety of quantization effects on averaging algorithms, and demonstrate favorable convergence properties by simulations. Our work adopts the Lyapunov method, as in [12], [16]; the common function used in these papers turns out, however, not to be a valid Lyapunov function for our averaging algorithm. This is due again to that the state sum does not remain invariant, and the augmented surplus evolution must also be taken into account. According to these features, we establish an appropriate Lyapunov function, and prove that bounding its decay time can be reduced to finding the hitting time in a certain Markov chain.

## A. Setup and Organization

Consider a digraph $\mathcal{G} = (\mathcal{V}, \mathcal{E})$, where $\mathcal{V} = \{1, ..., n\}$ is the node set, and $\mathcal{E} \subseteq \mathcal{V} \times \mathcal{V}$ the edge set. Each directed edge $(j, i)$ in $\mathcal{E}$, pointing from $j$ to $i$, denotes that agent $j$ communicates to agent $i$ (namely, the information flow is from $j$ to $i$). Selfloop edges are not allowed, i.e., $(i, i) \notin \mathcal{E}$. Communication among the nodes is by means of *gossip*: At each time instant, exactly one edge $(j, i) \in \mathcal{E}$ is activated independently from all earlier instants and with a time-invariant positive probability $p_{ji} \in (0, 1)$ such that $\sum_{(j,i) \in \mathcal{E}} p_{ji} = 1$.

To model the quantization effect in information flow, we assume that at time $k \in \mathbb{Z}_+$ (nonnegative integers), each node has an integer-valued state $x_i(k) \in \mathbb{Z}$, $i \in \mathcal{V}$; the aggregate state is denoted by $x(k) = [x_1(k) \cdots x_n(k)]^T \in \mathbb{Z}^n$. Let

$$\mathcal{X} := \{x : m \leq x_i \leq M, \ i \in \mathcal{V}\}, \tag{1}$$

for some (finite) constants $m, M$. Suppose throughout the paper that the initial state satisfies $x(0) \in \mathcal{X}$. Also, let $\mathbf{1} = [1 \cdots 1]^T$ be the vector of all ones.

For the convergence time analysis, we will impose the following two assumptions on the graph topology and the probability distribution of activating edges. Let $|\cdot|$ denote the cardinality of a set.

*Assumption* 1. The digraph $\mathcal{G}$ is *complete* (i.e., every node is connected to every other node by a directed edge). It follows that there are $|\mathcal{E}| = n(n-1)$ edges.

*Assumption* 2. The probability distribution on edge activation is *uniform*; namely, each edge can be activated with the same probability $p := 1/|\mathcal{E}|$.

The rest of this paper is organized as follows. In Section II, we formulate and solve the problem of convergence time analysis for the consensus algorithm. Then in Sections III and IV, we derive an upper bound for the convergence time of the averaging algorithm. Further, we compare convergence rates through a numerical example in Section V, and finally we state our conclusions in Section VI.

## II. CONVERGENCE TIME OF QUANTIZED CONSENSUS ALGORITHM

### A. Problem Formulation

First we recall the quantized consensus (**QC**) algorithm from [7]. Suppose that the edge $(j, i) \in \mathcal{E}$ is randomly activated at time $k$. Along the edge node $j$ sends to $i$ its state information, $x_j(k)$, but does not perform any update, i.e., $x_j(k+1) = x_j(k)$. On the other hand, node $i$ receives $j$'s state $x_j(k)$ and updates its own as follows:

(**R1**)  If $x_i(k) = x_j(k)$, then $x_i(k+1) = x_i(k)$;

(**R2**)  if $x_i(k) < x_j(k)$, then $x_i(k+1) \in (x_i(k), x_j(k)]$;

(**R3**)  if $x_i(k) > x_j(k)$, then $x_i(k+1) \in [x_j(k), x_i(k))$.

Let the subset $\mathscr{C}$ of $\mathbb{Z}^n$ be the set of general consensus states:

$$\mathscr{C} := \{x : x_1 = \cdots = x_n\}. \tag{2}$$

We say that the nodes achieve general consensus almost surely if for every initial state $x(0)$, there exist $T < \infty$ and $x^* \in \mathscr{C}$ such that $x(k) = x^*$ for all $k \geq T$ with probability one. Under **QC** algorithm, a necessary and sufficient graphical condition that ensures almost sure general consensus is that the digraph $\mathcal{G}$ contains a *globally reachable node* (i.e., a node that is connected to every other node via a directed path) [7]. Clearly if $\mathcal{G}$ is complete, then every node is globally reachable.

The convergence time of **QC** algorithm is the random variable $T_{qc}$ defined by $T_{qc} := \inf\{k \geq 0 : x(k) \in \mathscr{C}\}$. The mean convergence time (with respect to the probability distribution on edge activation), starting from a state $x_0 \in \mathcal{X}$, is then given by

$$E_{qc}(x_0) := \mathbb{E}\left[T_{qc} | x(0) = x_0\right]. \tag{3}$$

*Problem* 1. Let Assumptions 1 and 2 hold. Find an upper bound of the mean convergence time $E_{qc}(x_0)$ of **QC** algorithm with respect to all possible initial states $x_0 \in \mathcal{X}$.

We now present the main result of this section: an upper bound of the mean convergence time $E_{qc}(x_0)$ for all possible initial states $x_0 \in \mathcal{X}$.

*Theorem* 1. Let Assumptions 1 and 2 hold. Then

$$\max_{x_0 \in \mathcal{X}} E_{qc}(x_0) < n(n-1)(M-m) = O(n^2).$$

To derive this bound, we first provide preliminaries on the hitting time in finite Markov chains.

*B. Preliminaries on Hitting Time*

Let $\{X_k\}_{k\geq 0}$ be a Markov chain with a finite state space $\mathcal{S}$ and a transition probability matrix $P = (P_{ij})$ (e.g., [18]). The entry $P_{ij}$ denotes the one-step transition probability from state $i$ to state $j$. In particular, the diagonal entry $P_{ii}$ denotes the *selfloop* transition probability. A state $i \in \mathcal{S}$ is said to be *absorbing* if $P_{ii} = 1$. For a given $\{X_k\}_{k\geq 0}$, the *hitting time* of a subset $\mathcal{T}$ of $\mathcal{S}$ is the random variable $H_{\mathcal{T}}(\{X_k\}_{k\geq 0})$ defined by

$$H_{\mathcal{T}}(\{X_k\}_{k\geq 0}) := \inf\{l \geq 0 : X_l \in \mathcal{T}\}.$$

The mean time (with respect to the probability distribution specified by $P$) taken for the chain, starting from a state $i \in \mathcal{S}$, to hit $\mathcal{T}$ is given by

$$E_i := \mathbb{E}\left[H_{\mathcal{T}}(\{X_k\}_{k\geq 0}) | X_0 = i\right] = \sum_{l=0}^{\infty} l \cdot \mathbb{P}\left[H_{\mathcal{T}}(\{X_k\}_{k\geq 0}) = l | X_0 = i\right], \quad (4)$$

where $\mathbb{E}[\cdot|\cdot]$ and $\mathbb{P}[\cdot|\cdot]$ denote the conditional expectation and conditional probability operators, respectively. Here is an important fact on mean hitting times [18, Theorem 1.3.5].

*Lemma* 1. The vector of mean hitting times $(E_i)_{i\in\mathcal{S}}$ of a subset $\mathcal{T}$ satisfies the system of linear equations

$$\begin{cases} E_i = 0 & \text{for } i \in \mathcal{T}, \\ E_i = \sum_{j\notin\mathcal{T}} P_{ij} E_j + 1 & \text{for } i \notin \mathcal{T}. \end{cases}$$

Using Lemma 1, we derive a closed-form expression of the mean hitting time for a specific Markov chain; this chain will be shown to characterize the state transition structure under **QC** algorithm. For the proof of this result, see Appendix.

*Lemma* 2. Consider the Markov chain in Fig. 1 with transition probabilities

$$p_z + r_z + q_z = 1, \ p_z = q_z \ (z = 1, ..., n-1), \quad r_0 = 1, \quad r_n = 1.$$

Then the mean hitting time of the state 0 or $n$ starting from state $z$ is

$$E_z = (1 - \frac{z}{n})\sum_{i=1}^{z-1}\frac{i}{p_i} + \frac{z}{n}\sum_{j=z}^{n-1}\frac{n-j}{p_j} \ (z = 1, ..., n-1).$$

*C. Analysis of Convergence Time*

We now proceed as follows. For an arbitrary $x(k)$ define the minimum and maximum states by

$$m(k) := \min_{i\in\mathcal{V}} x_i(k), \ M(k) := \max_{i\in\mathcal{V}} x_i(k). \quad (5)$$

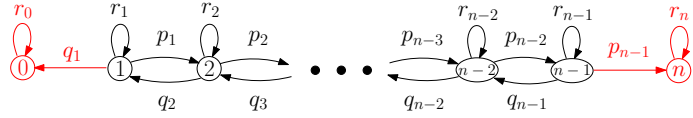

Fig. 1. Markov chain I: states 0 and $n$ are absorbing. Here $r_0, \ldots, r_n$ are selfloop transition probabilities.

We view the state $x(k)$ converging to $\mathscr{C}$ as the interval $[m(k), M(k)]$ shrinking to length 0. Let the random variable $T_{qc}^1$ be the time when one interval shrinkage occurs; then the corresponding mean time, starting from a state $x$, is $E_{qc}^1(x) := \mathbb{E}\left[T_{qc}^1 | x \in \mathcal{X}\right]$. Since one shrinkage decreases the interval length by at least 1, there can be at most $M - m$ shrinkages for $x_0 \in \mathcal{X}$. It then follows that

$$\max_{x_0 \in \mathcal{X}} E_{qc}(x_0) \leq \max_{x \in \mathcal{X}} E_{qc}^1(x) \cdot (M - m). \tag{6}$$

Consider a subset $\mathcal{X}_1$ of $\mathcal{X}$ defined by

$$\mathcal{X}_1 := \{x : x_1 = \cdots = x_z = 1 \ \& \ x_{z+1} = \cdots = x_n = 0, \ z \in [1, n-1]\}. \tag{7}$$

Thus the interval has length 1 for all $x \in \mathcal{X}_1$. It is not difficult to see that $\max_{x_0 \in \mathcal{X}_1} E_{qc}(x_0) = \max_{x \in \mathcal{X}} E_{qc}^1(x)$. The following lemma states an upper bound of $E_{qc}(x_0)$ for $x_0 \in \mathcal{X}_1$.

*Lemma* 3. Let Assumptions 1 and 2 hold. Then $\max_{x_0 \in \mathcal{X}_1} E_{qc}(x_0) < n(n-1) = O(n^2)$.

*Proof.* By Assumptions 1 and 2, every directed edge in $\mathcal{G}$ can be activated with the uniform probability $p = 1/(n(n-1))$. Starting from an arbitrary state in the set $\mathcal{X}_1$, the transition structure under **QC** algorithm is the Markov chain displayed in Fig. 1; in the diagram,

$$\begin{cases} \text{state } 0: & \text{the vector } \mathbf{0} = [0 \cdots 0]^T \text{ of all zeros,} \\ \text{state } n: & \text{the vector } \mathbf{1} = [1 \cdots 1]^T \text{ of all ones,} \\ \text{state } z: & \text{the vector } [\overbrace{1 \cdots 1}^{z} \ 0 \cdots 0]^T \text{ in } \mathcal{X}_1, \end{cases} \tag{8}$$

and the transition probabilities are $p_z = q_z = z(n-z)p$, $z \in [1, n-1]$. To see this, consider the transition from state $z$ to state $z+1$; this occurs when an edge $(j, i)$ is activated, with $x_j = 1$ and $x_i = 0$, so that **(R2)** of **QC** algorithm applies. Since there are $z(n-z)$ such edges, the transition probability $p_z = z(n-z)p$. Likewise, one may derive that the transition from state $z$ to state $z-1$ is with probability $q_z = z(n-z)p$, which occurs when **(R3)** of **QC** algorithm applies. Now observe in Fig. 1 that the states $0, n \in \mathscr{C}$ and $1, \ldots, n-1 \in \mathcal{X}_1$; hence $\max_{z \in [1, n-1]} E_z = \max_{x_0 \in \mathcal{X}_1} E_{qc}(x_0)$, where $E_z$ is from (4).

It is left to invoke the formula of $E_z$ in Lemma 2 for the obtained transition probabilities, which yields

$$E_z = (1 - \frac{z}{n}) \sum_{i=1}^{z-1} \frac{1}{(n-i)p} + \frac{z}{n} \sum_{j=z}^{n-1} \frac{1}{jp}$$

$$\leq (1 - \frac{z}{n}) \frac{z-1}{(n-z+1)p} + \frac{z}{n} \frac{n-z}{zp}$$

$$= \frac{n-z}{n-z+1} \cdot \frac{1}{p} < \frac{1}{p} = n(n-1).$$

Thus $E_z < n(n-1)$ for all $z \in [1, n-1]$. Therefore $\max_{x_0 \in \mathcal{X}_1} E_{qc}(x_0) < n(n-1) = O(n^2)$. ■

Finally, our main result (Theorem 1) on upper bounding $E_{qc}(x_0)$ for $x_0 \in \mathcal{X}$ follows immediately from Lemma 3 and (6).

*Remark* 1. We discuss the idea of how this result for complete graphs might be extended to handle more general topologies. We still view reaching consensus as the interval $[m(k), M(k)]$ shrinking to length 0; thereby the inequality (6) holds. We then again consider the subset $\mathcal{X}_1$ given in (7), and as long as the digraph is strongly connected (i.e., every node is connected to every other node) one can verify that the state transition structure under **QC** algorithm is still the one in Fig. 1. The associated transition probabilities, however, depend crucially on topologies. In order to apply again Lemma 2 to derive bounds, it would be important to establish the relation between transition probabilities and graph topologies; this will be targeted in our future work.

## III. QUANTIZED AVERAGING ALGORITHM AND ITS LYAPUNOV FUNCTION

In this and next sections, we address the convergence time analysis for the quantized averaging (**QA**) algorithm, which is a modification of the one in [7], [8]. We start by presenting the modified algorithm, and formulate the corresponding time analysis problem. We then propose a Lyapunov function, which turns out to be a suitable measure for the average consensus error. In Section IV, we will derive an upper bound on the mean convergence time by means of bounding the decay time of the proposed Lyapunov function.

### A. Problem Formulation

First we present **QA** algorithm. As in [7], [8], since the state sum cannot be preserved at each time instant, we associate each node $i \in \mathcal{V}$ with an additional *surplus* variable, $s_i(k) \in \mathbb{Z}$, to locally record the changes of $x_i(k)$. The aggregate surplus is denoted by $s(k) = [s_1(k) \cdots s_n(k)]^T \in \mathbb{Z}^n$, whose initial value is set to be $s(0) = [0 \cdots 0]^T$. Now suppose that the edge $(j, i) \in \mathcal{E}$ is activated at time $k$. There are two stages: (I) Along the edge, node $j$ sends to $i$ its state $x_j(k)$ and surplus $s_j(k)$. Node $j$ does not update its state, but sets its surplus to be 0 after transmission (see Fig. 2). (II) Based on the information received, node $i$ determines either to update its state and surplus, or to send back to $j$ the surplus $s_j(k)$ by activating the opposite edge $(i, j)$ (see Fig. 3). Notice that the latter operation in (II) requires bidirectional communication between two nodes at a single time instant; this is possible in complete digraphs (our assumption), but not in general strongly connected digraphs.

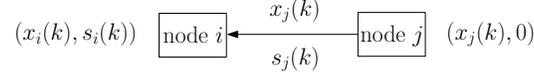

Fig. 2. Stage (I): Node $j$ sends to $i$ its state and surplus through the edge $(j, i)$.

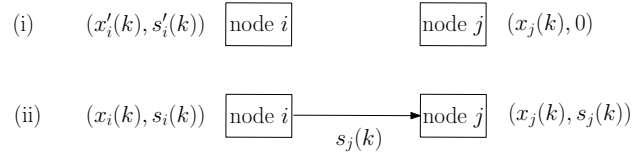

Fig. 3. Stage (II): Either (i) node $i$ updates its state and surplus, or (ii) it sends $s_j(k)$ back to node $j$ through edge $(i, j)$.

Formally, **QA** algorithm is described as follows.

**(R1)** If $x_i(k) = x_j(k)$, then there are two cases:

**(i)** If $s_i(k) > 0$ & $s_j(k) > 0$, then

$$x_i(k+1) = x_i(k), \quad s_i(k+1) = s_i(k);$$
$$x_j(k+1) = x_j(k), \quad s_j(k+1) = s_j(k).$$

**(ii)** Otherwise (i.e., either surplus equals zero),

$$x_i(k+1) = x_i(k), \quad s_i(k+1) = s_i(k) + s_j(k) \in \{0, 1\};$$
$$x_j(k+1) = x_j(k), \quad s_j(k+1) = 0.$$

**(R2)** If $x_i(k) < x_j(k)$, then there are two cases:

**(i)** If $s_i(k) + s_j(k) > 0$, then

$$x_i(k+1) = x_i(k) + 1, \quad s_i(k+1) = s_i(k) + s_j(k) - 1 \in \{0, 1\};$$
$$x_j(k+1) = x_j(k), \quad s_j(k+1) = 0.$$

**(ii)** Otherwise (i.e., $s_i(k) + s_j(k) = 0$),

$$x_i(k+1) = x_i(k), \quad s_i(k+1) = s_i(k) + s_j(k) = 0;$$
$$x_j(k+1) = x_j(k), \quad s_j(k+1) = 0.$$

**(R3)** If $x_i(k) > x_j(k)$, then there are two cases:

**(i)** If $s_i(k) + s_j(k) = 0$, then

$$x_i(k+1) = x_i(k) - 1, \quad s_i(k+1) = s_i(k) + s_j(k) + 1 = 1;$$
$$x_j(k+1) = x_j(k), \quad s_j(k+1) = 0.$$

**(ii)** Otherwise (i.e., $s_i(k) + s_j(k) > 0$),

$$x_i(k+1) = x_i(k), \quad s_i(k+1) = s_i(k);$$
$$x_j(k+1) = x_j(k), \quad s_j(k+1) = s_j(k).$$

In the algorithm, observe that (1) **(R1)(i)** and **(R3)(ii)** are where node $i$ sends $s_j(k)$ back to node $j$ in stage (II), which requires bidirectional communication; (2) only **(R3)(i)** 'generates' one surplus, and only **(R2)(i)** 'consumes' one surplus; (3) the quantity $(x+s)^T \mathbf{1}$ stays invariant, i.e., for every $k \geq 0$,

$$(x(k+1) + s(k+1))^T \mathbf{1} = (x(k) + s(k))^T \mathbf{1} = x(0)^T \mathbf{1}. \tag{9}$$

Distinct from the algorithm in [7], [8], this **QA** algorithm does not involve the threshold constant and the local extrema variables, thus reducing individual computation effort. Also each surplus variable is indeed binary-valued, and therefore requires merely one bit for both storage and transmission. A further difference between the two algorithms lies in the use of surplus variables: The algorithm in [7], [8] allows surpluses to pile up, which is indeed required to achieve average consensus for arbitrary strongly connected digraphs. By contrast, our **QA** algorithm here prevents surpluses from piling up, and meanwhile simplifies the transition structure. In addition, unlike the algorithm in [12] which assumes bidirectional communication for all time, the design of surplus updates here marks a feature of our **QA** algorithm.

Now let the subset $\mathscr{A}$ of $\mathbb{Z}^n \times \mathbb{Z}^n$ be the set of average consensus states:

$$\mathscr{A} := \{(x,s) : x_i = \lfloor x(0)^T \mathbf{1}/n \rfloor \text{ or } \lceil x(0)^T \mathbf{1}/n \rceil, \ i \in \mathcal{V}\}. \tag{10}$$

We say that the nodes achieve average consensus almost surely if for every initial condition $(x(0), 0)$, there exist $T < \infty$ and $(x^*, s^*) \in \mathscr{A}$ such that $(x(k), s(k)) = (x^*, s^*)$ for all $k \geq T$ with probability one. Here is the convergence result of **QA** algorithm for complete digraphs.

*Proposition* 1. Let Assumption 1 hold. Then, under **QA** algorithm, the nodes achieve average consensus almost surely.

This convergence result may be justified by a similar argument as given in [7], [8]; some care, however, has to be taken for the operations on surplus variables, as pointed out above. For completeness, the proof is provided in the Appendix. In addition, we note that the convergence can also be implied by the time analysis using Lyapunov approach in Section IV below.

The convergence time of **QA** algorithm is the random variable $T_{qa}$ defined by $T_{qa} := \inf\{k \geq 0 : (x(k), s(k)) \in \mathscr{A}\}$. The mean time taken for this convergence (according again to the probability distribution on edge activation), starting from $(x_0, 0)$ with $x_0 \in \mathcal{X}$, is then given by

$$E_{qa}(x_0) := \mathbb{E}\left[T_{qa} | (x(0), 0) = (x_0, 0)\right]. \tag{11}$$

*Problem* 2. Let Assumptions 1 and 2 hold. Find an upper bound of the mean convergence time $E_{qa}(x_0)$ of **QA** algorithm with respect to all possible initial states $x_0 \in \mathcal{X}$.

Our main result is the following upper bound of $E_{qa}(x_0)$ with respect to all possible initial states $x_0 \in \mathcal{X}$.

*Theorem* 2. Let Assumptions 1 and 2 hold. Then

$$\max_{x_0 \in \mathcal{X}} E_{qa}(x_0) < n^2(n-1)\frac{3(M-m)}{2} + n(n-1)\frac{R(R-1)}{n-(R/2)} = O(n^3),$$

where $R \in [0, n-1]$ is an integer, as in (12) below.

We note that the order of this polynomial bound is the same as that in [12] for undirected, complete graphs. To derive this bound, we will first propose a valid Lyapunov function for **QA** algorithm. Then we will upper bound the mean convergence time by way of upper bounding the mean decay time of the Lyapunov function.

*B. Lyapunov Function*

We start by introducing two variables, called positive surplus $S_+$ and negative surplus $S_-$; they are global variables, but are needed only for the convergence time analysis. Write the initial state sum

$$x(0)^T \mathbf{1} = nL + R, \tag{12}$$

where $L := \lfloor x(0)^T \mathbf{1}/n \rfloor$ is one of the possible values for average consensus, and $0 \leq R < n$. Observe that when a surplus is generated/consumed, the corresponding state moves one-step either closer to or farther from the value $L$. Positive and negative surplus variables are used to identify these two directions.

Concretely, when a surplus is generated, we increase $S_+$ (resp. $S_-$) if the corresponding state moves towards (resp. away from) $L$. On the other hand, when a surplus is consumed, we distinguish the following two situations: In one case where the state moves closer to $L$, we decrease $S_-$ if it is nonzero, and $S_+$ otherwise; in the other case where the state moves away from $L$, we decrease only $S_+$.

We now formalize the updating rules of $S_+$ and $S_-$. Let $D(k) := \sum_{i=1}^{n} |x_i(k) - L|$ be the sum of average consensus errors, and suppose that the edge $(j, i) \in \mathcal{E}$ is activated at time $k$.

**(S1)** If **(R3)(i)** generates one surplus, then there are two cases:

(i) If $D(k+1) = D(k) - 1$ (i.e., $x_i(k) > L$), then

$$S_+(k+1) = S_+(k) + 1.$$

(ii) If $D(k+1) = D(k) + 1$ (i.e., $x_i(k) \leq L$), then

$$S_-(k+1) = S_-(k) + 1.$$

**(S2)** If **(R2)(i)** consumes one surplus, then there are also two cases:

(i) If $D(k+1) = D(k) + 1$ (i.e., $x_i(k) \geq L$), then

$$S_+(k+1) = S_+(k) - 1.$$

(ii) If $D(k+1) = D(k) - 1$ (i.e., $x_i(k) < L$), then

$$S_-(k) = 0 \Rightarrow S_+(k+1) = S_+(k) - 1;$$
$$S_-(k) > 0 \Rightarrow S_-(k+1) = S_-(k) - 1.$$

**(S3)** Otherwise

$$S_+(k+1) = S_+(k);$$
$$S_-(k+1) = S_-(k).$$

The case **(S3)** above includes **(R1)**, **(R2)(ii)**, and **(R3)(ii)** of **QA** algorithm; note that, in these cases, there is no state update. Since initially there is no surplus in the system (i.e., $s(0) = 0$), we set $S_+(0) = S_-(0) = 0$. Also, one may readily see that $S_+(k) + S_-(k) = s(k)^T \mathbf{1}$, which relates the global surpluses to the local ones.

We are ready to define the Lyapunov function $V(k)$, $k \geq 0$, which is given by

$$V(k) := D(k) + S_+(k) - S_-(k). \tag{13}$$

It is not difficult to see from **(S1)**-**(S3)** that $V(k)$ is non-increasing. Indeed, $V(k)$ stays put except for only one case – **(S2)(ii)** and negative surplus $S_-(k) = 0$ – where it decreases by 2, i.e., $V(k+1) = V(k) - 2$. Notice that after this decrement, $S_+(k+1) \geq 0$ and $S_-(k+1) = 0$.

*Remark* 2. We emphasize that the validity of $V(k)$ as a Lyapunov function is not restricted only to undirected graphs, since the updating rules **(S2)** and **(S3)** do not involve **(R1)(i)** and **(R3)(ii)** where bidirectional communication is required. Indeed, $V(k)$ is a suitable Lyapunov function for the original **QA** algorithm in [7], [8], which can achieve average consensus on arbitrary strongly connected digraphs. This is one contribution of our work, which might also provide a preliminary to attack convergence time on more general topologies.

In the following lemma, we collect several useful implications from the definition of function $V(k)$.

*Lemma* 4.

(1) A lower bound of $V(k)$ is $R$, i.e., $V(k) \geq R$ for all $k$.

(2) If $V(k) = R$, then $S_-(k) = 0$, $S_+(k) \geq 0$, and $(\forall i \in [1, n])\ x_i(k) \geq L$.

(3) If $D(k) = 0$, then $S_-(k) = 0$ and $V(k) = S_+(k) = R$.

(4) Suppose $R = 0$. Then $D(k) = 0$ if and only if $V(k) = 0$, and in both cases $S_-(k) = S_+(k) = 0$.

*Proof.* We prove these statements in this order: (2), (1), (3), and (4).

(2) Let $V(k) = R$. Then there must exist $k_0 \leq k$ such that $V(k_0 - 1) = R + 2$ and $V(k_0) = R$. Also we have $S_+(k_0) \geq 0$ and $S_-(k_0) = 0$. Now assume $x_1(k_0) < L$. It follows from (9) that $x_1(k_0) + \sum_{i=2}^{n} x_i(k_0) + s(k_0)^T \mathbf{1} = nL + R$. Rearranging the terms and by $s(k_0)^T \mathbf{1} = S_+(k_0) + S_-(k_0)$, we obtain $\sum_{i=2}^{n} x_i(k_0) - (n-1)L = (L - x_1(k_0)) + R - S_+(k_0)$. Thus

$$V(k_0) = (L - x_1(k_0)) + \sum_{i=2}^{n} x_i(k_0) + S_+(k_0) - S_-(k_0)$$

$$= 2(L - x_1(k_0)) + R > R.$$

This contradicts $V(k_0) = R$, and hence $x_i(k_0) \geq L$ for all $i$. The latter holds also for time $k$ because the minimum states are non-decreasing by **QA** algorithm. Finally, according to the updating rules of $S_+$ and $S_-$, one may easily see that $S_-(k) = 0$ and $S_+(k) \geq 0$.

(1) When $V(k) = R$, every state $x_i(k) \geq L$ and consequently **(S3)(ii)** cannot occur. As $V(k)$ is non-increasing, it is lower bounded by $R$.

(3) Let $D(k) = 0$. Then $x(k)^T \mathbf{1} = nL$, and thus $S_+(k) + S_-(k) = s(k)^T \mathbf{1} = R$. It follows that $V(k) = S_+(k) - S_-(k) \leq R$. But $V(k) \geq R$, so that necessarily $V(k) = S_+(k) - S_-(k) = R$, which also implies that $S_-(k) = 0$ and $S_+(k) = R$.

(4) Assume $R = 0$. (Only if) The conclusion follows immediately from (3). (If) Let $V(k) = 0$. Then there must exist $k_0 \leq k$ such that $V(k_0 - 1) = 2$ and $V(k_0) = 0$. Also we have $S_+(k_0) \geq 0$ and $S_-(k_0) = 0$. Hence $D(k_0) + S_+(k_0) = 0$, which results in $D(k_0) = S_+(k_0) = 0$. As average consensus is achieved at $k_0$, no further state or surplus update occurs. So the conclusion for time $k$ follows. ■

Next, we find an upper bound for the function $V(k)$.

*Proposition 2.* Let $x(0) \in \mathcal{X}$ in (1). Then for every $k \geq 0$,
$$V(k) \leq \frac{(M-m)n}{2} + R.$$

*Proof.* Since the function $V(k)$ is non-increasing, it suffices to find an upper bound for $V(0) = \sum_{i=1}^n |x_i(0) - L|$. Consider the function $V(0) - R$; it is convex in $x(0)$, and $\mathcal{X}$ is a convex set. Hence, one of the extreme points of $\mathcal{X}$ is a maximizer. Fix $r \in [1, n]$, and let $x(0) \in \mathcal{X}$ be such that $x_1(0) = \cdots = x_r(0) = m$ and $x_{r+1}(0) = \cdots = x_n(0) = M$. Then $V(0) - R = r(L - m) + (n - r)(M - L) - R$. Also we have $L = (\mathbf{1}^T x(0) - R)/n = (rm + (n-r)M - R)/n$. Substituting this into the above equation and rearranging the terms, we derive

$$\begin{aligned}
V(0) - R &= -\frac{2(M-m)}{n}r^2 + \left(2(M-m) - 2\frac{R}{n}\right)r \\
&= \frac{2(M-m)}{n}\left[-\left(r - \frac{1}{2}(n - \frac{R}{M-m})\right)^2 + \frac{1}{4}(n - \frac{R}{M-m})^2\right] \\
&\leq \frac{2(M-m)}{n} \cdot \frac{1}{4}(n - \frac{R}{M-m})^2 \qquad \text{( equality holds iff } r = \frac{1}{2}(n - \frac{R}{M-m}) \text{ )} \\
&= \frac{1}{2}\frac{(n(M-m) - R)^2}{n(M-m)} \\
&\leq \frac{1}{2}\frac{(n(M-m))^2}{n(M-m)} = \frac{(M-m)n}{2} \qquad \text{( equality holds iff } R = 0 \text{ )}.
\end{aligned}$$

Thus $V(k) - R$ is upper bounded by $(M - m)n/2$, which is achievable if and only if $R = 0$ and $r = n/2$. ■

## IV. CONVERGENCE TIME ANALYSIS OF QA ALGORITHM

We turn now to analyzing the mean convergence time of **QA** algorithm, by way of upper bounding the mean decay time of the Lyapunov function $V(\cdot)$ in (13). This Lyapunov approach is also adopted in [12], [16]; the common function used is $V'(k) = \sum_{i=1}^n (x_i(k) - x(0)^T \mathbf{1}/n)^2$. It can be verified that $V'(k)$ is, however, not a valid Lyapunov function with respect to our **QA** algorithm. This is due to that the state sum is not preserved in each iteration and the surplus evolution must also be taken into account, as in our function $V(k)$.

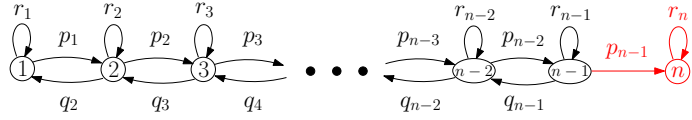

Fig. 4. Markov chain II: state $n$ is absorbing.

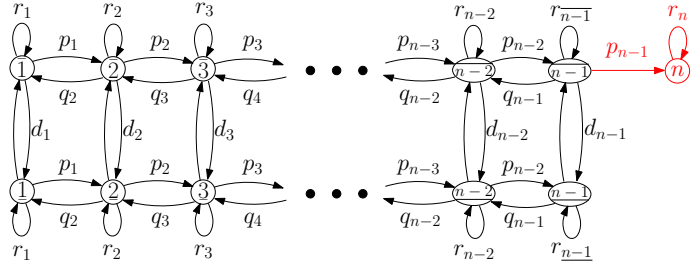

Fig. 5. Markov chain III: state $n$ is absorbing.

## A. Preliminaries on Hitting Time

As in Subsection II-B, we provide preliminaries on the hitting time in finite Markov chains, specific to the analysis of **QA** algorithm. For the proofs see Appendix.

*Lemma* 5. Consider the Markov chain in Fig. 4 with transition probabilities

$$p_1 + r_1 = 1, \qquad p_z + r_z + q_z = 1 \ (z = 2, ..., n-1), \qquad r_n = 1.$$

Then the mean hitting times of the state $n$ starting from state $1$ and $z$ are respectively

$$E_1 = \sum_{l=2}^{n-1} \left[ \left( \prod_{i=2}^{l} \frac{q_i}{p_i} \right) \cdot \frac{1}{p_1} + \sum_{j=2}^{l} \left( \prod_{i=j+1}^{l} \frac{q_i}{p_i} \right) \cdot \frac{1}{p_j} \right] + \frac{1}{p_1},$$

$$E_z = \sum_{l=z}^{n-1} \left[ \left( \prod_{i=2}^{l} \frac{q_i}{p_i} \right) \cdot \frac{1}{p_1} + \sum_{j=2}^{l} \left( \prod_{i=j+1}^{l} \frac{q_i}{p_i} \right) \cdot \frac{1}{p_j} \right] \quad (z = 2, ..., n-1).$$

*Lemma* 6. Consider the Markov chain in Fig. 5 with transition probabilities

$$p_1 + r_1 + d_1 = 1, \qquad p_z + r_z + q_z + d_z = 1 \ (z = 2, ..., n-2),$$

$$r_{\underline{n-1}} + q_{n-1} + d_{n-1} = 1, \qquad p_{n-1} + r_{\overline{n-1}} + q_{n-1} + d_{n-1} = 1, \qquad r_n = 1.$$

Here $\underline{\cdot}$ and $\overline{\cdot}$ denote the states of the lower and upper rows, respectively. Then for states $\underline{n-1}$ and $\overline{n-1}$,

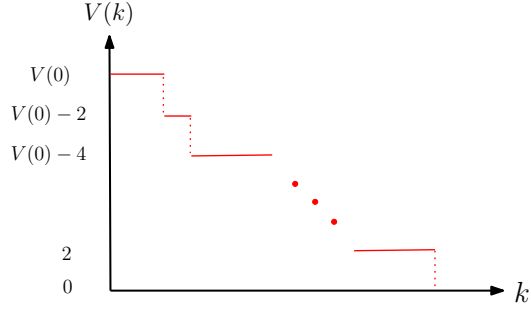

Fig. 6. Decay of function $V(k)$ in case $R = 0$

their mean hitting times of the absorbing state $n$ are

$$E_{\overline{n-1}} = \left(\prod_{i=2}^{n-1} \frac{q_i}{p_i}\right) \cdot \frac{2}{p_1} + \sum_{j=2}^{n-1} \left(\prod_{i=j+1}^{n-1} \frac{q_i}{p_i}\right) \cdot \frac{2}{p_j},$$

$$E_{\underline{n-1}} < \left(1 + \frac{p_{n-1}}{d_{n-1}}\right) E_{\overline{n-1}}.$$

In the rest of this section, the proof of Theorem 2 is given. We will need the following notation. Define the random variable $T_V := \inf\{k \geq 0 : V(k) = R\}$; thus $T_V$ is the time when $V(\cdot)$ decreases to $R$. The mean decay time, starting from $(x_0, 0)$, is then given by

$$E_V(x_0) := \mathbb{E}\left[T_V | (x(0), 0) = (x_0, 0)\right]. \tag{14}$$

Now recall $R$ from equation (12); we proceed with two cases in this order: $R = 0$ and $R > 0$. When $R = 0$ the mean convergence time $E_{qa}(x_0)$ is found to satisfy $E_{qa}(x_0) = E_V(x_0)$, whereas when $R > 0$ we have $E_{qa}(x_0) \geq E_V(x_0)$ in general and the corresponding analysis turns out to be based on the former case.

### B. Proof for the case $R = 0$

In this case, the mean convergence time $E_{qa}(x_0)$ is characterized by the mean time that the function $V(k)$ decays to 0; that is, $E_{qa}(x_0) = E_V(x_0)$ in (14). This is because by Lemma 4 (4), $V(k) = 0$ if and only if $D(k) = 0$, and the latter implies $(x(k), s(k)) \in \mathscr{A}$. As each decrement reduces $V(k)$ by 2, the initial value $V(0)$ is necessarily even, and there need in total $V(0)/2$ decrements.

To upper bound $E_V(x_0)$, we view the decay of $V(k)$ as the descent of *level sets* in the $(n+2)$-dimensional space of the triples $u := (x, S_+, S_-)$ (see Fig. 7). In this space, the average consensus state

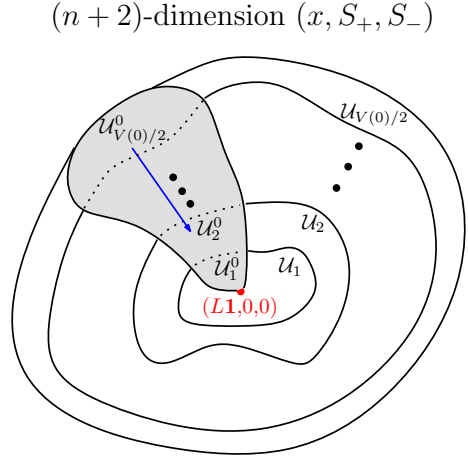

Fig. 7. Decay of $V(k)$ viewed as level set descent in the $(n+2)$ dimensions of $(x, S_+, S_-)$. Descending is possible only from the shaded area and through the dotted curves.

is simply the point $(L\mathbf{1}, 0, 0)$. Define the level sets

$$\mathcal{U}_l := \{u : V = \sum_{i=1}^{n} |x_i - L| + S_+ - S_- = 2 \cdot l\}, \quad l = 1, ..., V(0)/2.$$

Thus when $u(k) \in \mathcal{U}_l$, we interpret that $(x(k), s(k))$ is $l$-step away from $\mathscr{A}$ (i.e., $V(k)$ requires $l$ decrements to reach $0$). Also, it is important to note that on every level set $\mathcal{U}_l$, the triple evolution may start, and may descend to the next level, only from a strict subset $\mathcal{U}_l^0$ defined by

$$\mathcal{U}_l^0 := \{u \in \mathcal{U}_l : S_- = 0 \ \& \ S_+ \geq 0\}.$$

To see this, first recall that the decrement of $V(\cdot)$ (i.e., level set descent) requires $S_- = 0$ and $S_+ > 0$. Moreover, for the outmost level $\mathcal{U}_{V(0)/2}$, the initial triple is of the form $(x_0, 0, 0)$; and for each subsequent level, the triple evolution starts right after descending from the preceding level, where we have $S_- = 0$ and $S_+ \geq 0$.

Now let the random variable $T_1$ be the time of *one decrement* of $V(\cdot)$. The corresponding mean time, starting from a triple $u \in \mathcal{U}_l^0$, is then given by $E_1^l(u) := \mathbb{E}\left[T_1 | u \in \mathcal{U}_l^0\right]$, $l \in [1, V(0)/2]$. Since the initial value $V(0)$ is upper bounded by $(M-m)n/2$ (Proposition 2), the function $V(\cdot)$ requires at most $(M-m)n/4$ decrements to reach $0$. Hence, an upper bound of its mean decay time is the following:

$$\max_{x_0 \in \mathcal{X}} E_V(x_0) \leq \max_{l \in [1, V(0)/2], u \in \mathcal{U}_l^0} E_1^l(u) \cdot \frac{(M-m)n}{4}. \tag{15}$$

Here is a key result.

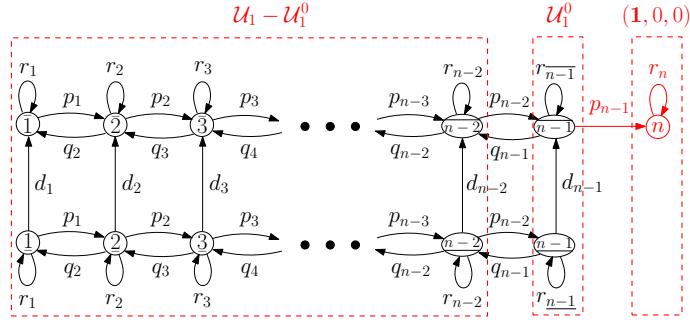

Fig. 8. One step away: from $\mathcal{U}_1$ to $(\mathbf{1}, 0, 0)$.

*Proposition* 3. Let Assumptions 1 and 2 hold. Then
$$\max_{l \in [1, V(0)/2], u \in \mathcal{U}_l^0} E_1^l(u) < 6n(n-1) = O(n^2).$$

To prove Proposition 3, it suffices to establish
$$\max_{u \in \mathcal{U}_l^0} E_1^l(u) < 6n(n-1) = O(n^2), \tag{16}$$

for every $l \in [1, V(0)/2]$. In the sequel we will provide the proof for the case $l = 1$ (i.e., one step away from average consensus), which contains the essential idea of our argument. Specifically, we first exhaust the possible triple evolution under **QA** algorithm, second derive the evolution structure and transition probabilities, and third calculate the corresponding mean hitting time. The analysis of the case $l \geq 2$ follows in a similar fashion but is more involved; we refer to Appendix for the proof.

*Proof for the case $l = 1$:* Without loss of generality let $L = 1$. We investigate the triple evolution from the level set $\mathcal{U}_1$, starting in $\mathcal{U}_1^0$, to the average consensus state $(\mathbf{1}, 0, 0)$. By Assumptions 1 and 2, every directed edge in $\mathcal{G}$ can be activated with the uniform probability $p = 1/(n(n-1))$. Consider the triple $([2 \ \overbrace{1 \cdots 1}^{n-2} \ 0]^T, \ 0, \ 0) \in \mathcal{U}_1^0$; we show that either $S_-$ or $S_+$ can be generated. Case 1: an edge $(j, i)$ is activated, with $x_j = 0$ and $x_i = 1$. In this case, **(R3)(i)** of **QA** algorithm applies, and the resulting triple is $([2 \ \overbrace{1 \cdots 1}^{n-3} \ 0 \ 0]^T, \ 0, \ 1) \in \mathcal{U}_1 - \mathcal{U}_1^0$. There are $n - 2$ such edges; so the probability of this transition is $(n-2)p$. In fact, such transitions can continue until all the ones become zeros, generating in total $S_- = n - 2$. Case 2: an edge $(j, i)$ is activated, with $x_j = 0$ or $1$ and $x_i = 2$. Again **(R3)(i)** of **QA** algorithm applies, the resulting triple being $([\overbrace{1 \ 1 \cdots 1}^{n-1} \ 0]^T, \ 1, \ 0) \in \mathcal{U}_1^0$. This transition is with probability $(n-1)p$, since there are $n - 1$ such edges.

Now starting from the triple $([\overbrace{1\ 1\cdots 1}^{n-1}\ 0]^T,\ 1,\ 0)$, on one hand, we can have a similar process, as from $([2\ \overbrace{1\cdots 1}^{n-2}\ 0]^T,\ 0,\ 0)$ described above, generating in total $S_- = n-2$. On the other hand, observe that there is only one edge $(j,i)$ such that $x_j = 1$, $s_j = 1$, and $x_i = 0$, $s_i = 0$. If this edge is activated (with probability $p$), then **(R2)(i)** of **QA** algorithm applies, and the resulting triple is the average consensus state $(\mathbf{1}, 0, 0)$.

Based on the above descriptions, we derive that the transition structure from $\mathcal{U}_1$ to $(\mathbf{1}, 0, 0)$ under **QA** algorithm is the one displayed in Fig. 8.[1] In this diagram, the state $n$ is the average consensus state $(\mathbf{1}, 0, 0)$, and the other states belong to $\mathcal{U}_1$, listed below:

$$
\begin{array}{ll}
\underline{n-1}:\ ([2\ 1\ 1\ \cdots\ 1\ 1\ 0]^T,\ 0,\ 0) & \overline{n-1}:\ ([1\ 1\ 1\ \cdots\ 1\ 1\ 0]^T,\ 1,\ 0) \\
\underline{n-2}:\ ([2\ 1\ 1\ \cdots\ 1\ 0\ 0]^T,\ 0,\ 1) & \overline{n-2}:\ ([1\ 1\ 1\ \cdots\ 1\ 0\ 0]^T,\ 1,\ 1) \\
\quad\vdots & \quad\vdots \\
\underline{2}:\ ([2\ 1\ 0\ \cdots\ 0\ 0\ 0]^T,\ 0,\ n-3) & \overline{2}:\ ([1\ 1\ 0\ \cdots\ 0\ 0\ 0]^T,\ 1,\ n-3) \\
\underline{1}:\ ([2\ 0\ 0\ \cdots\ 0\ 0\ 0]^T,\ 0,\ n-2) & \overline{1}:\ ([1\ 0\ 0\ \cdots\ 0\ 0\ 0]^T,\ 1,\ n-2)
\end{array}
$$

Note that negative surplus is zero ($S_- = 0$) only in the states $\underline{n-1}$ and $\overline{n-1}$; hence these two triples are in $\mathcal{U}_1^0$. Also, one may verify that the transition probabilities are as follows:

$$p_1 = (n-2)p, \quad d_1 = p; \quad p_{n-1} = p, \quad q_{n-1} = (n-2)p, \quad d_{n-1} = (n-1)p;$$

$$p_z = (n-1-z)zp, \quad q_z = (z-1)p, \quad d_z = zp \quad (z = 2, ..., n-2).$$

To upper bound $E_1^1(u)$ for $u \in \mathcal{U}_1^0$, in Fig. 8 we add transitions from the state $\overline{z}$ to $\underline{z}$ with the probability $d_z$, $z \in [1, n-1]$, thereby increasing the probabilities of moving away from the average consensus state $n$. This modification leads us to the same structure displayed in Fig. 5; thus, we have $\max_{u \in \mathcal{U}_1^0} E_1^1(u) \le E_{\underline{n-1}}$, where $E_{\underline{n-1}}$ is given in (4).

It is left to calculate $E_{\underline{n-1}}$ with respect to the obtained transition probabilities. For this we invoke the formulas in Lemma 6. First,

$$\prod_{i=2}^{n-1} \frac{q_i}{p_i} = \frac{n-2}{1} \cdot \frac{n-3}{n-2} \cdot \frac{n-4}{2(n-3)} \cdots \frac{2}{(n-4)3} \cdot \frac{1}{(n-3)2} = \frac{1}{(n-3)!}.$$

---

[1] The transition structure in Fig. 8 is obtained with a minor modification from the original. For those triples in $\mathcal{U}_1 - \mathcal{U}_1^0$, we treat the following transitions from left to right as selfloops: For some node $i$ such that $x_i = 0$ and $s_i = 0$, its state $x_i$ increases by consuming one negative surplus (under **R2(i)** of **QA** algorithm). By treating such transitions as selfloops, only the probability of moving towards the average consensus state is reduced; so it can be verified that the mean hitting time derived from this structure is an upper bound of that from the original. We make such modifications in our analysis henceforth.

Similarly,
$$\prod_{i=3}^{n-1} \frac{q_i}{p_i} = \frac{2}{(n-4)!}, \ \prod_{i=4}^{n-1} \frac{q_i}{p_i} = \frac{3}{(n-5)!}, \ \cdots, \ \frac{q_{n-2}q_{n-1}}{p_{n-2}p_{n-1}} = n-3, \ \frac{q_{n-1}}{p_{n-1}} = n-2.$$

We then have
$$\begin{aligned} E_{\overline{n-1}} &= \left(\prod_{i=2}^{n-1} \frac{q_i}{p_i}\right) \cdot \frac{2}{p_1} + \sum_{j=2}^{n-1} \left(\prod_{i=j+1}^{n-1} \frac{q_i}{p_i}\right) \cdot \frac{2}{p_j} \\ &= \frac{1}{(n-3)!} \cdot \frac{2}{(n-2)p} + \frac{2}{(n-4)!} \cdot \frac{2}{(n-3)2p} + \cdots + (n-2) \cdot \frac{2}{(n-2)p} + \frac{2}{p} \\ &= \frac{2}{p} \cdot \left[\frac{1}{(n-2)!} + \frac{1}{(n-3)!} + \cdots + 1 + 1\right] \\ &< \frac{2}{p} \cdot 3 = 6n(n-1) = O(n^2). \end{aligned}$$

Finally, $E_{\underline{n-1}} < (1 + (p_{n-1}/d_{n-1})) E_{\overline{n-1}} = (1 + (p/((n-1)p))) \cdot 6n(n-1) < 6n(n-1) = O(n^2)$. ∎

Therefore, it follows from Proposition 3 and equation (15) that the upper bound of $E_{qa}(x_0)$ in Theorem 2 holds for the case $R = 0$.

## C. Proof for the case $R \in [1, n-1]$

When $R \neq 0$, we have $E_{qa}(x_0) \geq E_V(x_0)$ in general. This is because $V(k) = R$ does not generally imply $(x(k), s(k)) \in \mathscr{A}$, and even after $V(k)$ reaches its lower bound $R$ (Lemma 4 (1) and (2)), the pair $(x(k), s(k))$ may require extra time to reach $\mathscr{A}$. Define the level set $\mathcal{U}_R := \{u : V = \sum_{i=1}^{n} |x_i - L| + S_+ - S_- = R\}$; then the mean convergence time starting from a triple $u \in \mathcal{U}_R$ is given by $E_{qa}(u) := \mathbb{E}[T_{qa}|u \in \mathcal{U}_R]$. Also recall from (14) that $E_V(x_0)$, with $x_0 \in \mathcal{X}$ in (1), denotes the mean decay time of $V(k)$ to the lower bound $R$. From these we obtain the mean convergence time of **QA** algorithm

$$\max_{x_0 \in \mathcal{X}} E_{qa}(x_0) \leq \max_{x_0 \in \mathcal{X}} E_V(x_0) + \max_{u \in \mathcal{U}_R} E_{qa}(u). \tag{17}$$

In the sequel, we find upper bounds for $E_V(x_0)$ and $E_{qa}(u_R)$, respectively. First, as in the case $R = 0$, we have

$$\max_{x_0 \in \mathcal{X}} E_V(x_0) < n^2(n-1)\frac{3(M-m)}{2} = O(n^3). \tag{18}$$

This is due to the following reason. The function $V(k)$ decays from its initial value $V(0)$ to $R$, and each decrement reduces $V(k)$ by 2. It follows that $V(0) - R$ is necessarily even and there need in total $(V(0) - R)/2$ decrements. For $l \in [1, (V(0) - R)/2]$ recall that $E_1^l(u)$ denotes the mean time spent for one decrement of $V(k)$, starting from a triple $u \in \mathcal{U}_l^0$. Following Proposition 3, one may similarly derive that $\max_{l \in [1,(V(0)-R)/2], u \in \mathcal{U}_l^0} E_1^l(u) < 6n(n-1)$. Moreover, $V(0) - R \leq (M-m)n/2$ by Proposition 2; thus $V(k)$ requires at most $(M-m)n/4$ decrements to reach $R$. Therefore, $\max_{x_0 \in \mathcal{X}} E_V(x_0) \leq \max_{l \in [1,(V(0)-R)/2], u \in \mathcal{U}_l^0} E_1^l(u) \cdot (M-m)n/4 < n^2(n-1)3(M-m)/2 = O(n^3)$.

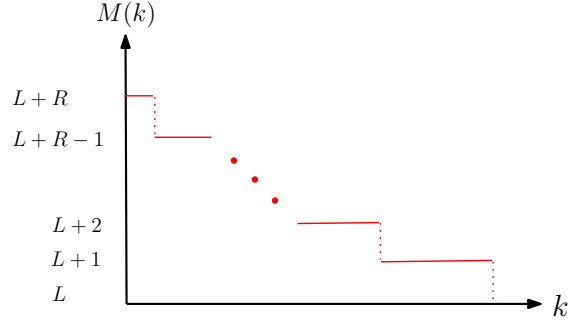

Fig. 9. Decrement of maximum state when $u \in \mathcal{U}_R$

Next, we find an upper bound for $\max_{u \in \mathcal{U}_R} E_{qa}(u)$. By Lemma 4 (2) we have $(\forall i \in \mathcal{V})\ x_i \geq L$; so the maximum state $M(k)$ in (5) satisfies $M(k) \in [L, L+R]$. If $R = 1$, then in fact $(x(k), s(k)) \in \mathscr{A}$; thus in this case $E_{qa}(u_R) = 0$, and we have from (17) and (18) that $\max_{x_0 \in \mathcal{X}} E_{qa}(x_0) = O(n^3)$. It is left to consider $R \in [2, n-1]$. Since $M(k) = L$ or $L+1$ implies $(x(k), s(k)) \in \mathscr{A}$, the mean convergence time $E_{qa}(u)$ can be characterized by the mean time that $M(k)$ decays to $L+1$. The decay of $M(k)$ is displayed in Fig. 9; observe that $M(k)$ requires at most $R-1$ decrements to reach $L+1$. Let $E_M(u)$ denote the mean time taken for one decrement of $M(k)$, starting from a triple $u \in \mathcal{U}_R$. Then an upper bound for $E_{qa}(u)$ is as follows:

$$\max_{u \in \mathcal{U}_R} E_{qa}(u) \leq \max_{u \in \mathcal{U}_R} E_M(u) \cdot (R-1). \tag{19}$$

*Proposition* 4. Let Assumptions 1 and 2 hold. Then

$$\max_{u \in \mathcal{U}_R} E_M(u) < n(n-1) \frac{R}{n - (R/2)} = O(n^2).$$

To prove Proposition 4, we first find the subset in which one decay of $M(k)$ takes the longest time, second derive the transition structure and probabilities under **QA** algorithm, and third compute the mean hitting time.

*Proof of Proposition 4.* We consider the following two cases when $R$ is even and odd, respectively.

1) $R$ is even. Let $\mathcal{U}_e$ be a subset of $\mathcal{U}_R$ given by $\mathcal{U}_e := \{u = (x, S_+, S_-) : x \in \mathcal{X}_e,\ S_+ = S_- = 0\}$, where

$$\mathcal{X}_e := \{x : x_1 = \cdots = x_{\frac{R}{2}} = L+2,\ x_{\frac{R}{2}+1} = \cdots = x_n = L\}.$$

For a state in $\mathcal{X}_e$, one decrement of its maximum value $L+2$ occurs only when all the $R/2$ state components having that value decrease; thus it is not hard to see $\max_{u\in\mathcal{U}_R} E_M(u) = \max_{u\in\mathcal{U}_e} E_M(u)$.

Now pick an arbitrary triple $u$ in $\mathcal{U}_e$; we investigate its evolution under **QA** algorithm. If an edge $(j,i)$ is activated, with $x_j = L$ and $x_i = L+2$, then **(R3)(i)** of **QA** algorithm applies, and the resulting triple is $([\overbrace{L+2\ \cdots\ L+2}^{(R/2)-1}\ L+1\ L\ \cdots\ L]^T,\ 1,\ 0)$. Namely, one maximum state decreases. Also observe that there are $(R/2)(n-(R/2))$ such edges; so the probability of this transition is $(R/2)(n-(R/2))p$, where $p = 1/(n(n-1))$ by Assumptions 1 and 2. Indeed, this process can continue until all the $R/2$ maximum states decrease to the value $L+1$, and we derive that the corresponding transition structure under **QA** algorithm is the one displayed in Fig. 4 with the length $n = (R/2)+1$. In the diagram,

$$\begin{cases} \text{state } 1: & ([\overbrace{L+2\ L+2\ \cdots\ L+2\ L+2}^{R/2}\ L\ \cdots\ L]^T,\ 0,\ 0) \\ \text{state } 2: & ([L+2\ L+2\ \cdots\ L+2\ L+1\ L\ \cdots\ L]^T,\ 1,\ 0) \\ \ \ \vdots & \\ \text{state } R/2: & ([L+2\ L+1\ L+1\ \cdots\ L+1\ L\ \cdots\ L]^T,\ (R/2)-1,\ 0) \\ \text{state } (R/2)+1: & ([L+1\ L+1\ L+1\ \cdots\ L+1\ L\ \cdots\ L]^T,\ R/2,\ 0) \end{cases}$$

and the transition probabilities are $p_1 = (R/2)(n-(R/2))p$, $p_z = ((R/2)-z+1)(n-(R/2))p$, $q_z = (z-1)((R/2)-z+1)p$, $z \in [2, R/2]$. Observe that the state $1 \in \tilde{\mathcal{U}}_R$ and the state $(R/2)+1 \in \mathcal{A}$; so $\max_{u\in\mathcal{U}_e} E_M(u) = E_1$, where $E_1$ is from (4).

It remains to invoke the formulas in Lemma 5 to calculate $E_1$. First,

$$\prod_{i=2}^{R/2} \frac{q_i}{p_i} = \frac{(R/2)-1}{n-(R/2)} \cdot \frac{((R/2)-2)2}{2(n-(R/2))} \cdots \frac{2((R/2)-2)}{((R/2)-2)(n-(R/2))} \cdot \frac{(R/2)-1}{((R/2)-1)(n-(R/2))}$$

$$= \frac{((R/2)-1)!}{(n-(R/2))^{(R/2)-1}} \leq \left(\frac{(R/2)-1}{n-(R/2)}\right)^{(R/2)-1} < 1;$$

the last inequality is due to $R < n$. Similarly $\prod_i^{R/2} q_i/p_i < 1$ for $i = 3, ..., R/2$. Then we obtain

$$\left(\prod_{i=2}^{l} \frac{q_i}{p_i}\right)\cdot\frac{1}{p_1} + \sum_{j=2}^{l}\left(\prod_{i=j+1}^{l}\frac{q_i}{p_i}\right)\cdot\frac{1}{p_j} < \frac{1}{(n-(R/2))p}\left(\frac{1}{(R/2)} + \frac{1}{(R/2)-1} + \cdots + \frac{1}{(R/2)-l+1}\right).$$

Hence,

$$E_1 = \sum_{l=2}^{R/2} \left[ \left( \prod_{i=2}^{l} \frac{q_i}{p_i} \right) \cdot \frac{1}{p_1} + \sum_{j=2}^{l} \left( \prod_{i=j+1}^{l} \frac{q_i}{p_i} \right) \cdot \frac{1}{p_j} \right] + \frac{1}{p_1}$$

$$< \frac{1}{(n-(R/2))p} \left( \frac{1}{R/2} + \frac{1}{(R/2)-1} + \cdots + \frac{1}{2} + 1 \right) + \frac{1}{(n-(R/2))p} \left( \frac{1}{R/2} + \frac{1}{(R/2)-1} + \cdots + \frac{1}{2} \right)$$

$$+ \cdots + \frac{1}{(n-(R/2))p} \left( \frac{1}{R/2} + \frac{1}{(R/2)-1} \right) + \frac{1}{(n-(R/2))p} \cdot \frac{1}{R/2}$$

$$= \frac{R}{(n-(R/2))p} = \frac{R}{(n-(R/2))} \cdot n(n-1).$$

Therefore, $\max_{u \in \mathcal{U}_R} E_M(u) = E_1 < n(n-1)R/(n-(R/2)) = O(n^2)$.

2) $R$ is odd. Let $\mathcal{U}_o$ be a subset of $\mathcal{U}_R$ given by $\mathcal{U}_o := \{u = (x, S_+, S_-) : x \in \mathcal{X}_o, \ S_+ = S_- = 0\}$, where

$$\mathcal{X}_o := \{x : x_1 = \cdots = x_{\frac{R-1}{2}} = L+2, \ x_{\frac{R+1}{2}} = L+1, \ x_{\frac{R+1}{2}+1} = \cdots = x_n = L\}.$$

For the same reason in the preceding case, one can verify that $\max_{u \in \mathcal{U}_R} E_M(u) = \max_{u \in \mathcal{U}_o} E_M(u)$. Also it turns out that the transition structure, together with the associated transition probabilities, starting from $\mathcal{U}_o$ is analogous to that starting from $\mathcal{U}_e$. Thus by a similar derivation given above, we can conclude again that $\max_{u \in \mathcal{U}_R} E_M(u) < n(n-1)R/(n-(R/2)) = O(n^2)$. ∎

Finally, it follows from equations (17)-(19) and Proposition 4 that an upper bound of the mean convergence time $E_{qa}(x_0)$ of **QA** algorithm is $E_{qa}(x_0) < n^2(n-1)3(M-m)/2 + n(n-1)R(R-1)/(n-(R/2)) = O(n^3)$ for the case $R > 0$. This completes the proof of Theorem 2.

*Remark* 3. We have derived an upper bound for the convergence time of **QA** algorithm on complete graphs, by proposing a suitable Lyapunov function for the algorithm and characterizing a Markov chain for the state-surplus transition structure. To extend this result to more general topologies, the Lyapunov function is still valid (see Remark 2) which in turn validates inequalities (15) and (17). Thus it is crucial to establish the relation between graph topologies and the transition structure with associated probabilities, as done in the proofs of Propositions 3 and 4 for complete graphs. Establishing such a relation for general topologies currently appears to be difficult, but will be explored in our future work.

## V. NUMERICAL EXAMPLE

We have proved polynomial upper bounds on the convergence time of **QC** and **QA** algorithms for complete digraphs. Now we compare these theoretic bounds with numerical simulations, so as to illustrate the tightness of our derived results. For this purpose, we consider the following initial states $x(0)$ which correspond to the worst case convergence time: For **QC** algorithm, we choose $x(0) = [\overbrace{1 \cdots 1}^{\lfloor n/2 \rfloor} \; 0 \cdots 0]^T$ (cf. proof of Lemma 3); for **QA** algorithm, we choose $x(0) = [2 \; \overbrace{1 \cdots 1}^{n-2} \; 0]^T$ (cf. proof of Proposition 3). The simulation results are displayed in Fig. 10, each plotted value being the mean convergence time of 100 runs of the corresponding algorithms.

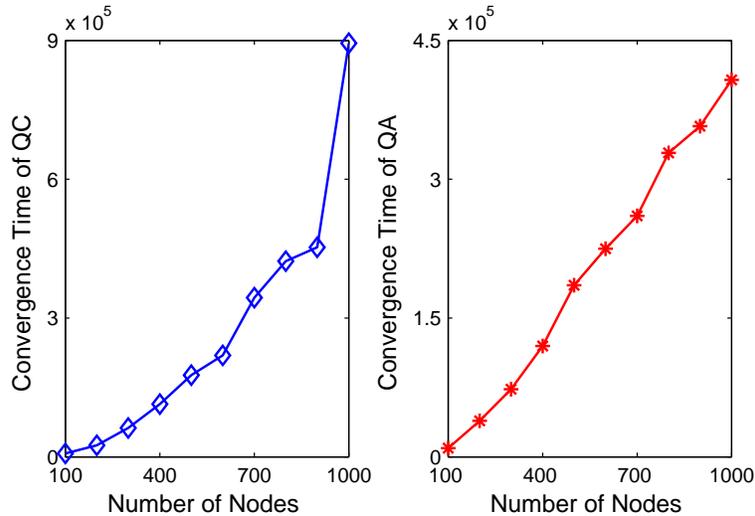

Fig. 10. Convergence time of **QC** and **QA**

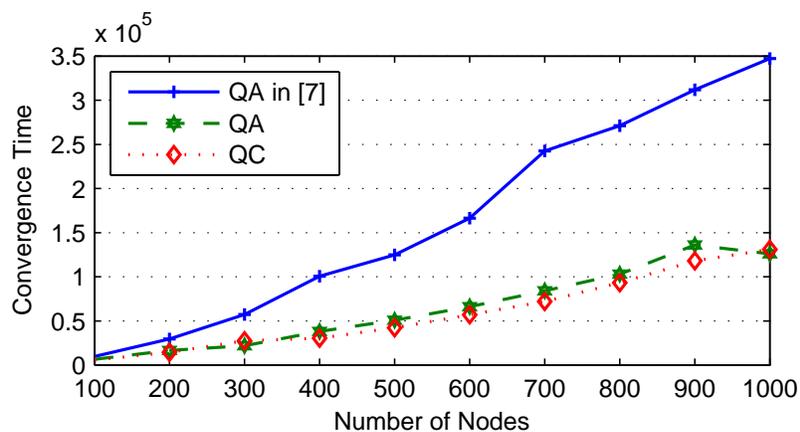

Fig. 11. Convergence time comparison among **QC**, **QA**, and **QA** in [7]

It is observed that the convergence rate of **QC** algorithm is approximately quadratic, which demonstrates that the derived theoretic bound is relatively tight. On the other hand, the convergence rate of **QA** algorithm appears to be at most quadratic, if not linear. This indicates that the cubic theoretic bound may not be tight, though it is in the same order as the one in [12] also for complete graphs. Thus, deriving tighter bounds for the convergence time of **QA** algorithm awaits future effort.

Furthermore, we compare the convergence rates of **QC**, **QA**, and the original **QA** algorithm in [7]. The results are shown in Fig. 11, each plotted value being the mean convergence time of 100 runs of the corresponding algorithms, with the initial states chosen uniformly at random from the interval $[-5, 5]$.

First it is observed that the convergence rates of **QC** and **QA** algorithms (dotted and dashed curves) are indeed analogous under the same initial conditions. Also we see that the **QA** algorithm in this paper is considerably faster than that in [7]. This improvement demonstrates that by occasionally requiring bidirectional communication, the modifications we have made for **QA** algorithm effectively accelerate convergence. This observation, on the other hand, indicates that there needs extra effort to bound the convergence time of the original **QA** algorithm in [7], which is for average consensus on general digraphs. This will be targeted in our future work.

## VI. CONCLUSIONS

In this paper, we have studied convergence time of the quantized gossip algorithms in [7], [8] which solve the consensus and averaging problems on digraphs. Specifically, we have derived upper bounds – polynomials in the number $n$ of nodes – on the mean convergence time of these algorithms for the special case of complete digraphs where the problem becomes tractable. For the consensus algorithm, the mean convergence time is $O(n^2)$; this is obtained by bounding the shrinking time of the smallest interval containing all states, which results in the special transition structure in Fig. 1. For the averaging algorithm, a valid Lyapunov function is proposed and its decay time investigated; this leads us to characterizing the convergence time by the hitting time in the Markov chains in Figs. 4 and 5, from which we derive $O(n^3)$ time complexity.

For future work, it would be of ample interest to analyze convergence time of our gossip algorithms on more general graph topologies, similar to the work of [13], [16]. A primary difficulty could lie in the potentially greater complexity of the state and surplus transition structure, resulting from the topological constraints. An alternative might be to explore the relation between the bounds for convergence time and the spectral properties of the Laplacian matrix associated to a given topology, as was done in [15].

## APPENDIX

*Proof of Lemma 2.* The proof is a direct calculation. By Lemma 1 the mean hitting times of state 0 or $n$ satisfy the following linear equations

$$E_0 = 0, \tag{20}$$

$$E_z = p_z E_{z+1} + r_z E_z + q_z E_{z-1} + 1, \quad z = 1, ..., n-1, \tag{21}$$

$$E_n = 0. \tag{22}$$

Since $p_z = q_z$, it follows from (21) that $p_z(E_{z+1} - E_z) - p_z(E_z - E_{z-1}) + 1 = 0$. Let $F_{z+1} := E_{z+1} - E_z$. Then

$$F_{z+1} = F_z - \frac{1}{p_z}.$$

This is a non-homogeneous first-order linear difference equation, whose solution is of the general form

$$F_{z+1} = F_1 - \sum_{i=1}^{z} \frac{1}{p_i}.$$

To obtain the initial condition $F_1$, consider

$$F_n + F_{n-1} + \cdots + F_1 = (E_n - E_{n-1}) + (E_{n-1} - E_{n-2}) + \cdots + (E_1 - E_0) = 0,$$

$$F_n + F_{n-1} + \cdots + F_1 = nF_1 - \sum_{j=1}^{n-1} \sum_{i=1}^{j} \frac{1}{p_i}.$$

From the above we have $F_1 = (1/n) \sum_{j=1}^{n-1} \sum_{i=1}^{j} 1/p_i$. Finally,

$$E_z = E_z - E_0 = F_z + F_{z-1} + \cdots + F_2 + F_1$$

$$= zF_1 - \sum_{j=1}^{z-1} \sum_{i=1}^{j} \frac{1}{p_i}$$

$$= \frac{z}{n} \sum_{j=1}^{n-1} \sum_{i=1}^{j} \frac{1}{p_i} - \sum_{j=1}^{z-1} \sum_{i=1}^{j} \frac{1}{p_i}$$

$$= (1 - \frac{z}{n}) \sum_{i=1}^{z-1} \frac{i}{p_i} + \frac{z}{n} \sum_{j=z}^{n-1} \frac{n-j}{p_j}.$$

∎

*Proof of Lemma 5.* By Lemma 1 the mean hitting times of state $n$ satisfy the following linear equations

$$E_n = 0, \tag{23}$$

$$E_1 = p_1 E_2 + r_1 E_1 + 1, \tag{24}$$

$$E_z = p_z E_{z+1} + r_z E_z + q_z E_{z-1} + 1, \quad z = 2, ..., n-1. \tag{25}$$

Rearrange the terms in (25) to obtain $p_z(E_{z+1} - E_z) - q_z(E_z - E_{z-1}) + 1 = 0$. Let $F_{z+1} := E_{z+1} - E_z$. Then

$$F_{z+1} = \frac{q_z}{p_z} F_z - \frac{1}{p_z},$$

whose initial condition is $F_2 = E_2 - E_1 = -1/p_1$ by (24). This is a non-homogeneous first-order linear difference equation with variable coefficients, whose solution is of the general form

$$F_{z+1} = \left( \prod_{i=2}^{z} \frac{q_i}{p_i} \right) \cdot \left( -\frac{1}{p_1} \right) + \sum_{j=2}^{z} \left( \prod_{i=j+1}^{z} \frac{q_i}{p_i} \right) \cdot \left( -\frac{1}{p_j} \right).$$

Since

$$F_n + F_{n-1} + \cdots + F_{z+1} = (E_n - E_{n-1}) + (E_{n-1} - E_{n-2}) + \cdots + (E_{z+1} - E_z)$$
$$= E_n - E_z = -E_z,$$

we derive

$$E_z = -(F_n + F_{n-1} + \cdots + F_{z+1}) = \sum_{l=z}^{n-1} \left[ \left( \prod_{i=2}^{l} \frac{q_i}{p_i} \right) \cdot \frac{1}{p_1} + \sum_{j=2}^{l} \left( \prod_{i=j+1}^{l} \frac{q_i}{p_i} \right) \cdot \frac{1}{p_j} \right].$$

Finally,

$$E_1 = E_2 + \frac{1}{p_1} = \sum_{l=2}^{n-1} \left[ \left( \prod_{i=2}^{l} \frac{q_i}{p_i} \right) \cdot \frac{1}{p_1} + \sum_{j=2}^{l} \left( \prod_{i=j+1}^{l} \frac{q_i}{p_i} \right) \cdot \frac{1}{p_j} \right] + \frac{1}{p_1}.$$

∎

*Proof of Lemma 6.* It follows from Lemma 1 that the mean hitting times of state $n$ satisfy the following linear equations

$$\begin{cases} E_{\overline{1}} = p_1 E_{\overline{2}} + r_1 E_{\overline{1}} + d_1 E_{\underline{1}} + 1, \\ E_{\underline{1}} = p_1 E_{\underline{2}} + r_1 E_{\underline{1}} + d_1 E_{\overline{1}} + 1; \end{cases} \quad (26)$$

$$\begin{cases} E_{\overline{z}} = p_z E_{\overline{z+1}} + r_1 E_{\overline{z}} + q_z E_{\overline{z-1}} + d_z E_{\underline{z}} + 1, \\ E_{\underline{z}} = p_z E_{\underline{z+1}} + r_1 E_{\underline{z}} + q_z E_{\underline{z-1}} + d_z E_{\overline{z}} + 1; \end{cases} \quad (z = 2, ..., n-2) \quad (27)$$

$$\begin{cases} E_{\overline{n-1}} = p_{n-1} E_n + r_{\overline{n-1}} E_{\overline{n-1}} + q_{n-1} E_{\overline{n-2}} + d_{n-1} E_{\underline{n-1}} + 1, \\ E_{\underline{n-1}} = r_{\underline{n-1}} E_{\underline{n-1}} + q_{n-1} E_{\underline{n-2}} + d_{n-1} E_{\overline{n-1}} + 1; \end{cases} \quad (28)$$

$$E_n = 0. \quad (29)$$

Rearrange the terms in (27) as

$$\begin{cases} p_z(E_{\overline{z+1}} - E_{\overline{z}}) - q_z(E_{\overline{z}} - E_{\overline{z-1}}) - d_z((E_{\overline{z}} - E_{\underline{z}})) + 1 = 0, \\ p_z(E_{\underline{z+1}} - E_{\underline{z}}) - q_z(E_{\underline{z}} - E_{\underline{z-1}}) + d_z((E_{\overline{z}} - E_{\underline{z}})) + 1 = 0. \end{cases}$$

Let $F_{\overline{z+1}} := E_{\overline{z+1}} - E_{\overline{z}}$, $F_{\underline{z+1}} := E_{\underline{z+1}} - E_{\underline{z}}$, and add these two equations; we obtain

$$F_{\overline{z+1}} + F_{\underline{z+1}} = \frac{q_z}{p_z}\left(F_{\overline{z}} + F_{\underline{z}}\right) - \frac{2}{p_z},$$

whose initial condition is $F_{\overline{2}} + F_{\underline{2}} = -2/p_1$ by (26). This is again a non-homogeneous first-order linear difference equation with variable coefficients, whose solution is

$$F_{\overline{z+1}} + F_{\underline{z+1}} = \left(\prod_{i=2}^{z} \frac{q_i}{p_i}\right) \cdot \left(-\frac{2}{p_1}\right) + \sum_{j=2}^{z} \left(\prod_{i=j+1}^{z} \frac{q_i}{p_i}\right) \cdot \left(-\frac{2}{p_j}\right).$$

Now rearrange the terms in (28)

$$\begin{cases} p_{n-1}(E_n - E_{\overline{n-1}}) - q_{n-1}(E_{\overline{n-1}} - E_{\overline{n-2}}) - d_{n-1}((E_{\overline{n-1}} - E_{\underline{n-1}})) + 1 = 0, \\ -q_{n-1}(E_{\underline{n-1}} - E_{\underline{n-2}}) + d_{n-1}((E_{\overline{n-1}} - E_{\underline{n-1}})) + 1 = 0. \end{cases}$$

Adding these two equations and applying (29), we derive

$$E_{\overline{n-1}} = -\frac{q_{n-1}}{p_{n-1}}\left(F_{\overline{n-1}} + F_{\underline{n-1}}\right) + \frac{2}{p_{n-1}} = \left(\prod_{i=2}^{n-1} \frac{q_i}{p_i}\right) \cdot \frac{2}{p_1} + \sum_{j=2}^{n-1}\left(\prod_{i=j+1}^{n-1} \frac{q_i}{p_i}\right) \cdot \frac{2}{p_j}.$$

It is left to obtain the upper bound for $E_{\underline{n-1}}$. For this we start by rearranging the terms in (26) as follows:

$$\begin{cases} (p_1 + d_1)E_{\overline{1}} - d_1 E_{\underline{1}} = p_1 E_{\overline{2}} + 1, \\ (p_1 + d_1)E_{\underline{1}} - d_1 E_{\overline{1}} = p_1 E_{\underline{2}} + 1. \end{cases}$$

Subtracting the first equation from the second, we have $(p_1 + 2d_1)(E_{\underline{1}} - E_{\overline{1}}) = p_1(E_{\underline{2}} - E_{\overline{2}})$. Hence

$$E_{\underline{1}} - E_{\overline{1}} = \frac{p_1}{p_1 + 2d_1}(E_{\underline{2}} - E_{\overline{2}}) < E_{\underline{2}} - E_{\overline{2}}.$$

Similarly, from (27) we obtain a chain of inequalities

$$E_{\underline{2}} - E_{\overline{2}} < E_{\underline{3}} - E_{\overline{3}} < \cdots < E_{\underline{n-2}} - E_{\overline{n-2}} < E_{\underline{n-1}} - E_{\overline{n-1}}.$$

Finally, rearrange the terms in (28) as

$$\begin{cases} (p_{n-1} + q_{n-1} + d_{n-1})E_{\overline{n-1}} - d_{n-1}E_{\underline{n-1}} = p_{n-1}E_n + q_{n-1}E_{\overline{n-2}} + 1, \\ (q_{n-1} + d_{n-1})E_{\underline{n-1}} - d_{n-1}E_{\overline{n-1}} = q_{n-1}E_{\underline{n-2}} + 1. \end{cases}$$

Subtracting the first equation from the second and applying (29), we deduce

$$(q_{n-1} + 2d_{n-1})(E_{\underline{n-1}} - E_{\overline{n-1}}) - p_{n-1}E_{\overline{n-1}} = q_{n-1}(E_{\underline{n-2}} - E_{\overline{n-2}}) < q_{n-1}(E_{\underline{n-1}} - E_{\overline{n-1}}).$$

Rearranging these terms we have $E_{\underline{n-1}} < (1 + (p_{n-1}/d_{n-1}))E_{\overline{n-1}}$. ∎

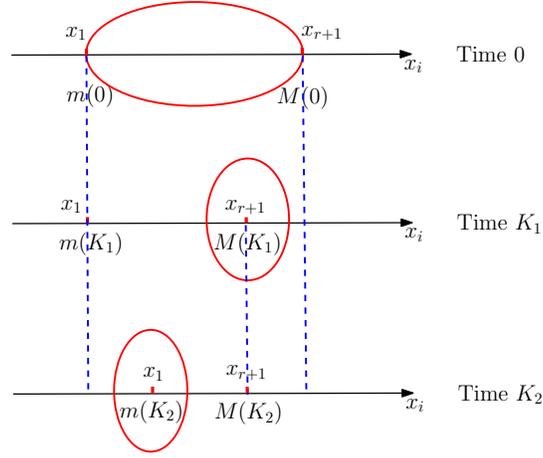

Fig. 12. Idea of induction step

*Proof of Proposition 1.* Based on [12, Theorem 2], it suffices to establish the following three conditions:

**(C1)** The evolution of $(x(k), s(k))$, $k \geq 0$, is a Markov chain with a finite state space;

**(C2)** the set $\mathscr{A}$ defined in (10) is an invariant set under **QA** algorithm;

**(C3)** for every $(x(0), 0) \notin \mathscr{A}$ there is a finite time $K_a$ such that $\Pr\big[(x(K_a), s(K_a)) \in \mathscr{A} \mid (x(0), 0)\big] > 0$.

For an arbitrary state $x(k)$, observe in **QA** algorithm that the minimum $m(k)$ is non-decreasing and the maximum $M(k)$ non-increasing, where $m(k)$, $M(k)$ are defined in (5). Thus the conditions **(C1)** and **(C2)** easily follow. It remains to establish **(C3)** when the digraph $\mathcal{G}$ is complete (Assumption 1), for which we proceed by induction on the number $n \, (> 1)$ of nodes. Let $F(k) := M(k) - m(k)$. Assume $(x(0), 0) \notin \mathscr{A}$; then $F(0) \geq 2$.

(i) Base case: $n = 2$. Label the two nodes such that $x_1(0) = m(0)$ and $x_2(0) = M(0)$. As $\mathcal{G}$ is complete, there are two edges, $(1, 2)$ and $(2, 1)$, each of which has a positive probability to be activated. Consider the sequence of alternate activation: $(1, 2), (2, 1), (1, 2), (2, 1) \cdots$. Then in **QA** algorithm, **(R3)(i)** and **(R2)(i)** will alternately apply, thereby shrinking the interval $[m(k), M(k)]$. It is easy to see that there exist a finite time $K_a$ and a positive probability such that $x_1(K_a) = x_2(K_a) = \lfloor (x_1(0) + x_2(0))/2 \rfloor$ (thus $(x(K_a), s(K_a)) \in \mathscr{A}$), and at most one node holds a surplus. Also in this process, $M(k)$ decreases by at least one and $m(k)$ increases by at least one.

(ii) Induction step: let $r \in [2, n-1]$. Suppose that for a network of $r$ nodes, there exist a finite time $K_a$ and a positive probability such that $x_1(K_a) = \cdots = x_r(K_a) = \lfloor (1/r) \sum_{i=1}^{r} x_i(0) \rfloor$, and at most $r - 1$ nodes each holds one surplus. Also suppose that in this process, $M(k)$ decreases by at least one and $m(k)$ increases by at least one.

Now consider the case with $r+1$ nodes. Label them such that $m(0) = x_1(0) \leq \cdots \leq x_{r+1}(0) = M(0)$. In the sequel, we describe a sequence of activating edges, which causes the interval $[m(k), M(k)]$ to shrink, the process being displayed in Fig. 12. The existence of the selected edges follows from that $\mathcal{G}$ is complete; and since each edge has a positive probability to be activated, the sequence of activation also has a positive probability.

First, consider the nodes $2, \ldots, r+1$. We distinguish three cases as follows.

Case 1: $x_{r+1}(0) - x_2(0) \geq 2$. Then applying the hypothesis, we obtain that in a finite time $K_1$ and with a positive probability, $x_2(K_1) = \cdots = x_{r+1}(K_1) = \lfloor (1/r) \sum_{i=2}^{r+1} x_i(0) \rfloor$.

Case 2: $x_{r+1}(0) - x_2(0) = 1$. For each node $i$ ($> 2$) such that $x_i(0) - x_2(0) = 1$, activate the edge $(2, i)$; then **(R3)(i)** of **QA** algorithm applies, thereby resulting again in $x_2(K_1) = \cdots = x_{r+1}(K_1) = \lfloor (1/r) \sum_{i=2}^{r+1} x_i(0) \rfloor$.

In both cases above, the maximum state decreases as $M(K_1) < M(0)$; hence $F(K_1) < F(0)$. In addition, there are at most $r-1$ nodes each having one surplus. Activate (one at a time, in an arbitrary order) the edges connecting those nodes with a surplus to the node 1. Thus **(R2)(i)** applies, and the surpluses are consumed to increase $x_1(k)$, which in turn causes $F(k)$ to decrease. At time at most $K_1' := K_1 + r - 1$, all the surpluses in the system can be consumed.

Case 3: $x_{r+1}(0) - x_2(0) = 0$. For this special case, we proceed directly to the next step.

Second, consider the nodes $1, \ldots, r$. When $F(K_1') \geq 2$ (or Case 3 above), applying the hypothesis we derive that in a finite time $K_2$ and with a positive probability, $x_1(K_2) = \cdots = x_r(K_2) = \lfloor (1/r) \sum_{i=1}^{r} x_i(K_1') \rfloor$. Since the minimum state $m(k)$ increases by at least one, we have $F(K_2) < F(K_1')$. Also, at most $r-1$ nodes each has one surplus. Select (one at a time, in an arbitrary order) the edges connecting the node $r+1$ to those with a surplus; then **(R2)(i)** applies, and the surpluses are consumed. Note that, however, here $F(k)$ stays put. At time at most $K_2' := K_2 + r - 1$, all the surpluses in the system can be consumed. If $F(K_2') \geq 2$, we apply the hypothesis again for the nodes $2, \ldots, r+1$, as is done in the first step above.

Thus we can repeat these two steps, in an alternate fashion, so that $F(k)$ decreases until $F(K_a') = 1$, for some finite time $K_a'$. There are two possibilities: (1) $x_1(K_a') = m(K_a')$, others $m(K_a')+1$, and at most $r-1$ nodes each has one surplus; and (2) $x_{r+1}(K_a') = M(K_a')$, others $M(K_a')-1$, and at most $r-1$ nodes each has one surplus. Analogous to the edge activation done above, one can show in both scenarios that there exist a finite time $K_a > K_a'$ and a positive probability such that $F(K_a) = 0$, and at most $r$ nodes each has one surplus. Therefore necessarily, $x_1(K_a) = \cdots = x_{r+1}(K_a) = \lfloor (1/(r+1)) \sum_{i=1}^{r+1} x_i(0) \rfloor$. Finally, it is evident that in this averaging process, $M(k)$ decreases by at least one and $m(k)$ increases by at least one. This finishes the induction step. ∎

*Proof of Proposition 3.* We have given in Section IV-B the proof for the case $l = 1$, one step away from average consensus. It remains to establish (16) for every $l \in [2, V(0)/2]$. Before proceeding, we introduce the following notation for an economical representation of the transition structure in Fig. 8:

$$([1\ 1\ 1\ \cdots\ 1\ 1\ 0]^T,\ 1,\ 0)$$
$$\uparrow$$
$$([2\ 1\ 1\ \cdots\ 1\ 1\ 0]^T,\ 0,\ 0)$$

Here $([1\ 1\ 1\ \cdots 1\ 1\ 0]^T,\ 1,\ 0)$ represents the upper row of states $\overline{1}, \ldots, \overline{n-1}$, and $([2\ 1\ 1\ \cdots 1\ 1\ 0]^T,\ 0,\ 0)$ represents the lower row of states $\underline{1}, \ldots, \underline{n-1}$. It is well to note that the state $n$ (i.e., the average consensus state $(\mathbf{1}, 0, 0)$) is not involved. Observe that only the triples in $\mathcal{U}_1^0$ are used, and only the triple with positive surplus $S_+ > 0$ has a transition probability to the average consensus state. We will use this notation to display the transition structures in the subsequent analysis.

(i) Two steps away: from $\mathcal{U}_2$ to $\mathcal{U}_1$. The corresponding transition structure is displayed in Fig. 13; there are four triples, representing four rows similar to the above. These rows can be arranged into three blocks $\mathcal{B}_1$, $\mathcal{B}_2$, and $\mathcal{B}_3$ as shown. Notice that the displayed triples are all in $\mathcal{U}_2^0$, and only those triples with positive surplus $S_+ > 0$ have a transition probability to $\mathcal{U}_1$. One can readily see that starting from the triple $([3\ 1\ 1 \cdots 1\ 1\ -1]^T, 0, 0)$, the mean hitting time of $\mathcal{U}_1$ is the longest; thus we need to analyze the whole structure.

In the sequel, the structure will be simplified in two steps. First, treat the transition to $\mathcal{B}_3$ as a selfloop at the triple $([2\ 1\ 1 \cdots 1\ 1\ -1]^T, 1, 0)$ in $\mathcal{B}_2$. This modification increases the mean hitting time starting from $\mathcal{B}_1$. To see this, note that the triple in $\mathcal{B}_3$ has more positive surplus $S_+$, which results in higher probabilities of moving towards $\mathcal{U}_1$. It then follows that selflooping in $\mathcal{B}_2$ takes longer time to hit $\mathcal{U}_1$ than transiting to $\mathcal{B}_3$. Second, combine $([3\ 1\ 1 \cdots 1\ 1\ -1]^T, 0, 0)$ in $\mathcal{B}_1$ and $([2\ 2\ 1 \cdots 1\ 1\ -1]^T, 0, 0)$ in $\mathcal{B}_2$. This amounts to combining the corresponding two rows of triples. It can be verified that the associated transition probabilities in these two rows are the same, except for those moving to $([2\ 1\ 1 \cdots 1\ 1\ -1]^T, 1, 0)$. Since the latter means moving towards $\mathcal{U}_1$, taking the smaller transition probabilities from the two rows will increases the mean hitting time.

After the above modifications, the transition structure is simplified to the one displayed in Fig. 5, with

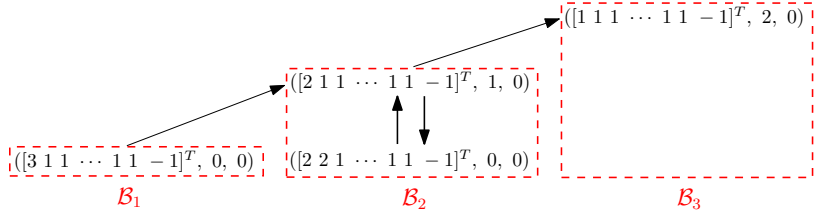

Fig. 13. Two steps away: from $\mathcal{U}_2$ to $\mathcal{U}_1$.

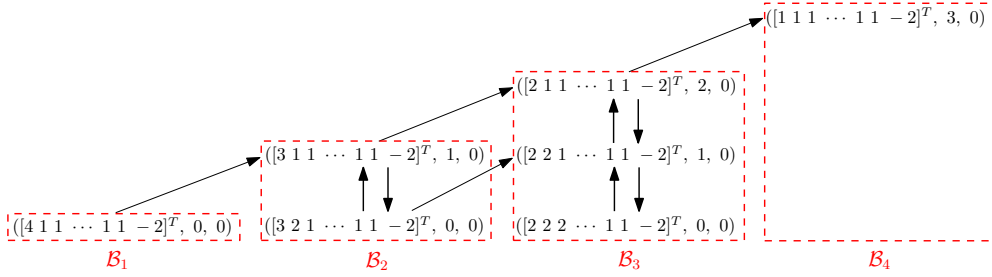

Fig. 14. Three steps away: from $\mathcal{U}_3$ to $\mathcal{U}_2$.

the following transition probabilities:

$$p_1 = (n-2)p, \quad d_1 = p; \quad p_{n-1} = p, \quad q_{n-1} = (n-2)p, \quad d_{n-1} = (n-2)p;$$

$$p_z = (n-1-z)zp, \quad q_z = (z-1)p, \quad d_z = (z-1)p \quad (z = 2, ..., n-2).$$

Hence, we have $\max_{u \in \mathcal{U}_2^0} E_2^1(u) \leq \underline{E_{n-1}}$, where $\underline{E_{n-1}}$ is given in (4). Invoke the formulas in Lemma 6, and perform an analogous calculation as before; we then obtain that $\max_{u \in \mathcal{U}_2^0} E_1^2(u) = O(n^2)$.

(ii) Three steps away: from $\mathcal{U}_3$ to $\mathcal{U}_2$. The corresponding transition structure is displayed in Fig. 14; we now have four blocks. Since starting from the triple $([4\ 1\ 1\cdots 1\ 1\ -2]^T, 0, 0)$ the mean hitting time of $\mathcal{U}_2$ is the longest, we need to analyze again the whole structure.

We take three steps to simplify the structure. First, treat the transition to $\mathcal{B}_4$ as a selfloop at the triple $([2\ 1\ 1\cdots 1\ 1\ -2]^T, 2, 0)$ in $\mathcal{B}_3$. This is the same as that in (ii), and hence increases the mean hitting time starting from $\mathcal{B}_1$. Second, treat the transitions to block $\mathcal{B}_3$ as selfloops at the corresponding triples in $\mathcal{B}_2$. This modification also increases the mean hitting time. To see this, compare the structure of $\mathcal{B}_2$ and its counterpart in $\mathcal{B}_3$ (i.e., the lower two triples alone). One may verify that the former has longer rows of triples and higher probabilities of moving away from $\mathcal{U}_2$. Hence, the mean time taken to hit $\mathcal{U}_2$ in the structure of $\mathcal{B}_2$ is longer than that in its counterpart in $\mathcal{B}_3$. Further, the top triple $([2\ 1\ 1\cdots 1\ 1\ -2]^T, 2, 0)$ in $\mathcal{B}_3$, with more positive surplus $S_+$, makes the mean hitting time even shorter. Therefore, selflooping in $\mathcal{B}_2$ increases the mean time to hit $\mathcal{U}_2$ compared to transiting to $\mathcal{B}_3$. Lastly, combine $([4\ 1\ 1\cdots 1\ 1\ -2]^T, 0, 0)$ in $\mathcal{B}_1$ and $([3\ 2\ 1\cdots 1\ 1\ -2]^T, 0, 0)$ in $\mathcal{B}_2$, as is done in (ii).

The above simplifications lead us again to the structure displayed in Fig. 5, with exactly the same transition probabilities as (ii). We thus obtain $\max_{u \in \mathcal{U}_3^0} E_3^1(u) \leq E_{\underline{n-1}} = O(n^2)$.

(iii) General $l\ (>3)$ steps away: from $\mathcal{U}_l$ to $\mathcal{U}_{l-1}$. The corresponding transition structure consists of $l+1$ blocks. Apply an analogous procedure to simplify this structure; it can be found by a similar argument that transiting to further blocks will accelerate hitting $\mathcal{U}_{l-1}$. Consequently, the structure with $l+1$ blocks can also be reduced to the one in Fig. 5, the probabilities of which are those in (ii). Therefore, $\max_{u \in \mathcal{U}_l^0} E_l^1(u) \leq E_{\underline{n-1}} = O(n^2)$. ∎